# Interfacial charge-induced adsorption mode for electron pairing in high-temperature superconductors

Jiu Hui Wu [1*], Hua Tian [1], and Kejiang Zhou [2]

[1]*School of Mechanical Engineering, Xi'an Jiaotong University,*

*& State Key Laboratory for Strength and Vibration of Mechanical Structures, Xi'an 710049, China*

[2] *Huzhou Institute of Zhejiang University, Huzhou 313000, China*

* E-mail: ejhwu@xjtu.edu.cn

**Abstract**: The electron pairing mechanism by the interfacial charge-induced adsorption mode of high-temperature superconductors is revealed. For the YBCO superconductors, the coupling of electrons and valence-flexible state of oxygen ions forms a charge-regulated interfacial layer induced by the adsorption potential, and electrons are paired by sharing the optimized interfacial structure and exchanging the adsorption mode, generating strong attraction to form Cooper pairs. Then the effective interaction potential between electrons is exactly derived in details, as well as the electron-adsorption-mode coupling strength, in which the adsorption coupling constant is up to 43.4. Furthermore, we verify that *d*-wave come from the anisotropy of interfacial adsorption forces, and explain the pseudo-energy gap behavior. By using the one-dimensional Ginzburg-Landau equation in the absence of a magnetic field, we obtain the coherence length expression, and the coherence length calculated is very close to the literature results. By establishing the energy gap equation, we obtain the superconducting gap $\Delta = 18.37\text{meV}$, which is very close to the measured result of 17meV by Scanning Tunneling Microscopy/Spectroscopy. These quantitative predictions close to the known results could verify our theoretical framework.

## 1. Introduction

The discovery of cuprate superconductors with transition temperatures exceeding liquid nitrogen temperatures [1,2] has posed one of the most persistent challenges in condensed matter physics. Despite over three decades of intensive research, the mechanism responsible for electron pairing in these materials remains controversial [3]. The Bardeen-Cooper-Schrieffer (BCS) theory [4], successful for conventional superconductors, appears insufficient for cuprates due to: (i) Transition temperatures (Tc~90-135 K) far exceeding the McMillan limit (~40 K) [5]; (ii) Anomalous d-wave pairing symmetry [6]; (iii) Pseudogap phenomena in the underdoped regime [7]; (iv) Non-Fermi liquid behavior in the normal state [8]; (v) Weak and non-monotonic isotope effects [9]; (vi) Extreme sensitivity to carrier concentration and layer structure [10].

These "anomalies" have motivated numerous theoretical proposals, including spin fluctuation-mediated pairing [11], resonating valence bond (RVB) states [12], and quantum criticality scenarios [13]. However, no consensus has emerged, and the field continues to seek a unifying principle.

Over the past five years, research on high-temperature superconductivity (HTS) mechanisms has achieved breakthroughs in nickelate superconductors, universal scaling in cuprate/iron-based systems, pseudogap and Higgs mode dynamics, and hydride high-Tc mechanisms. Nickel-based superconductors (nickelates) have emerged as the third unconventional HTS family (after cuprates and iron-based), with critical advances in ambient-pressure stabilization, pairing mechanisms, and new material discovery [14-17]. A pivotal breakthrough is the universal linear scaling between superfluid density ($\rho_{s0}$) and strange-metal resistivity coefficient ($A_1$) across iron-based and cuprate superconductors, proving a common quantum-critical fluctuation origin for both phenomena [18-19]. Clarifying the pseudogap nature and its interplay with superconductivity is a longstanding challenge. Recent advances use strong-field terahertz third-harmonic generation to detect Higgs mode dynamics and PG-superconductivity coupling [20-21]. High-pressure hydrides (e.g., $CaH_{14}$, $LaH_{10}$) exhibit near-room-temperature superconductivity (Tc ~ 200–288 K) under megabar pressure, with mechanisms rooted in strong e-ph coupling and pressure-induced nearly free electrons [22-23]. In addition, a unified theory integrated BCS theory, density functional theory, and Zentropy theory across cuprates, iron-based, and hydride superconductors has been built [24].

Wu et al. [25-26] proposed that superconducting electrons can be understood as adsorbed onto lattice surfaces via the thermodynamic potential. The molar adsorption potential $\varepsilon = A(\mu_s - \mu_n)$ was shown to correlate linearly with Tc across diverse superconductors, from elemental Sn to $HgBa_2Ca_2Cu_3O_{8+\delta}$. The same group [27-28] developed a quantum state equation based on Thom's catastrophe theory [29-30], introducing a phase transition index α that characterizes the crossover from quantum to classical behavior. For superconductors, $\alpha$ jumps from 15/14 (normal state) to 15/11 (superconducting state) at Tc [25].

For high-temperature superconducting cuprates, the necessary conditions for a high Tc can be summarized as follows: 1) The presence of $CuO_2$ planes; 2) An appropriate number of holes on the $CuO_2$ planes; 3) The holes on the $CuO_2$ planes are mobile. The consensus reached so far is that high-temperature superconductors are derived from heavily hole-doped oxides with strong valence flexibility.

In this paper, the forming mechanism of Cooper pairs in high-temperature superconductors is investigated, and our central thesis is that high-temperature superconductivity emerges from the interplay of (i) long-range interfacial adsorption potentials, (ii) electron pairing by sharing the optimized interfacial charge regulation induced by adsorption potential, and (iii) structurally stable catastrophe transitions. We synthesize these three strands into a coherent theoretical framework.

This paper is organized as follows: Section 2 introduces the interfacial charge regulation induced by the adsorption potential. Section 3 derives the effective interaction potential between electrons. Section 4 derives the superconducting gap equation, the coherence length, pairing symmetry, and their quantitative predictions. In Section 5 the conclusion is given.

## 2. Interfacial charge regulation induced by adsorption potential

Since the composition and layered structure of superconductors play a decisive role in the adsorption potential, the $CuO_2$ planes and Cu-O chain layers in $YBa_2Cu_3O_7$ exhibit distinctly different local adsorption potentials. The $CuO_2$ planes possess a stronger adsorption potential, which raises the local carrier concentration and drives carrier condensation and Cooper-pair formation required for superconductivity. In contrast, the Cu-O chain layers show a weaker adsorption potential and merely serve as a charge reservoir, where holes are localized and do not contribute to superconductivity.

In the $YBa_2Cu_3O_7$ superconductor cell, since the -2 valence state of oxygen is not complete, only with one hole located on the oxygen site, the total number of holes per unit cell is approximately 1.0, of which only about half transfer into the $CuO_2$ planes and act as effective superconducting carriers, while the other half are localized in the Cu-O chain layers as a charge reservoir. This asymmetric interlayer charge separation could be explained jointly by the thermodynamic adsorption potential and its induced interfacial charge regulation.

Therefore, the interlayer distribution of holes and the origin of the effective superconducting component in $YBa_2Cu_3O_7$ are attributed to the spatial modulation and selective condensation of carriers by the lattice-surface adsorption potential in the layered structure, i.e. the interlayer charge separation in $YBa_2Cu_3O_7$ is a consequence of layer-dependent adsorption potentials.

Furthermore, the effective superconducting component ratio $\omega$ is defined as the fraction of total carriers that contribute to Cooper-pair condensation. For the optimal $YBa_2Cu_3O_7$ superconductor (Tc=90 K), the calculated value is $\omega$=0.0305976, meaning only 3.06% of the total carriers participate in superconductivity [25].

This small but crucial fraction originates from the layer-dependent adsorption potential in the layered structure. The strong adsorption potential in the $CuO_2$ planes selectively drives carrier condensation and superconductivity, while the weak potential in the Cu-O chain layers localizes most holes as a charge reservoir. Thus, the interlayer separation of holes and the magnitude of $\omega$ are both determined by the spatial modulation of the thermodynamic adsorption potential.

We think that the effective superconducting carriers in $YBa_2Cu_3O_7$ arise from hole redistribution driven by the adsorption potential, and the coupling of electrons and valence-flexible state of oxygen ions forms a charge-regulated interfacial layer induced by the adsorption potential. Further the spontaneous valence flexibility of multi-hole carriers coupled with surface charges can store enormous interfacial free energy. When two electrons approach each other, this energy is released by sharing the optimized interfacial structure, generating strong attraction to form Cooper pairs.

## 3. Electron pairing by interfacial charge-induced adsorption modes

The electron gas near the lattice surface is localized within an adsorption space of a certain thickness under the constraint of the adsorption potential, forming a quantum confined system in the direction perpendicular to the surface. In the adsorption potential framework, the phonon-mediated mechanism of BCS will be replaced by the adsorption modes induced by the interfacial charge regulation, which describes the collective oscillation modes of the electron gas near the lattice surface. The adsorption mode is the intrinsic dynamical response of this confined system, while the adsorption potential serves as the macroscopic thermodynamic binding energy that maintains this quantum confined structure. Actually the adsorption modes are the microscopic dynamic manifestations of the adsorption potential, and the adsorption potential is the macroscopic thermodynamic integral of the adsorption modes.

For high-temperature superconducting copper oxides, electrons are confined in the adsorption space between the two layers of $CuO_2$ with a thickness of $d_{\mathrm{ads}} \sim 0.3\mathrm{nm}$, which is closer to a deep potential well confinement because $\hbar^2/(m^* {d_{\mathrm{ads}}}^2) \sim 100\mathrm{meV} \gg \varepsilon_p \sim 3\mathrm{meV}$, where ħ is Plank's constant and $m^*$ is the effective mass of electron. In this case, the adsorption mode frequency $\omega_{\mathrm{ads}} \sim 1/{d_{\mathrm{ads}}}^2$ is essentially a scale effect caused by the combined geometric dependence of the quantum confinement of the electron gas in the direction perpendicular to the surface and the electrostatic restoring force, which could also be obtained by the scaling relations in the quantum state equation [26].

In the adsorption potential theory, when an electron enters the adsorption space near the crystal surface, the effective potential it experiences is the chemical potential difference $\varepsilon_p \geq \mu_s - \mu_n$, where $\mu_s$ and $\mu_n$ are of the superconducting phase and normal phase respectively. The local electronic density fluctuation $\delta n(\mathbf{r})$ causes change in the local chemical potential, i.e. $\delta\mu(\mathbf{r}) = \frac{\partial\mu}{\partial n}\delta n(\mathbf{r})$ ($n$ is the number of electrons per unit volume), which is precisely the microscopic origin of the interaction between electrons and the adsorbed phase.

In the unified Hamiltonian, the coupling term between electrons and the lattice surface adsorption mode (adsorption potential fluctuations) is

$$\widehat{H}_{el-ads} = \int d^3r \delta\hat{\mu}(\boldsymbol{r})\hat{\rho}(\boldsymbol{r}) \tag{1}$$

where by box normalization with the volume $V$ the electron density operator is

$$\hat{\rho}(\boldsymbol{r}) = \frac{1}{V}\sum_{\boldsymbol{k},\boldsymbol{q}} \mathrm{e}^{-i\boldsymbol{q}\cdot\boldsymbol{r}} c^{\dagger}_{\boldsymbol{k}+\boldsymbol{q},\sigma} c_{\boldsymbol{k}\sigma} \tag{2}$$

and the chemical potential response $\delta\hat{\mu}(\boldsymbol{r})$ can be obtained by the Bogoliubov linearization approximation for the condensed uniform electron background $n_s$ (superconducting electron concentration) as

$$\delta\hat{\mu}(\boldsymbol{r}) = \frac{\partial\mu}{\partial n}\delta\hat{n}_{ads}(\boldsymbol{r}) = \frac{\partial\mu}{\partial n}\frac{1}{\sqrt{V}}\sum_{\boldsymbol{q}} \mathrm{e}^{i\boldsymbol{q}\cdot\boldsymbol{r}} \sqrt{n_s}\left(b_{\boldsymbol{q}} + b^{\dagger}_{-\boldsymbol{q}}\right) \tag{3}$$

where $c^{\dagger}_{\boldsymbol{k}\sigma}$ and $c_{\boldsymbol{k}\sigma}$ are, respectively, the creation and annihilation operators for an electron of spin σ, and $b_q$ is the absorption mode (Boson) annihilation operator.

Substituting Eqs. (2) and (3) into Eq. (1) and using $\int d^3r\, \mathrm{e}^{-i(\boldsymbol{q}-\boldsymbol{q}')\cdot\boldsymbol{r}} = V\delta_{\boldsymbol{q}\boldsymbol{q}'}$, there is

$$\widehat{H}_{el-ads} = \sum_{\boldsymbol{k},\boldsymbol{q}} \frac{\partial\mu}{\partial n}\sqrt{\frac{n_s}{V}}\, c^{\dagger}_{\boldsymbol{k}+\boldsymbol{q},\sigma}\, c_{\boldsymbol{k}\sigma}\left(b_{\boldsymbol{q}} + b^{\dagger}_{-\boldsymbol{q}}\right) \quad (4)$$

where $g_q^{ads} = \frac{\partial\mu}{\partial n}\sqrt{\frac{n_s}{V}}$ is the electron-adsorption-mode coupling strength determined by the adsorption potential $\varepsilon_p$ and adsorption spatial thickness $d_{\mathrm{ads}}$.

An effective interaction between two electrons is generated through the exchange of adsorbed modes. Similar to the derivation of phonon-induced attraction in BCS theory, the degrees of freedom of the adsorption mode are treated with second-order perturbation using the Schrieffer-Wolff transformation, diagonalizing the electron-adsorption-mode coupling Hamiltonian into the electronic subspace. Because the interaction is not instantaneous but involves a time delay caused by the propagation of the Boson mode (adsorption mode), the coupling between electrons and adsorbed modes is described using real-frequency delayed interactions. By eliminating the intermediate states of the collective adsorption modes, the effective interaction potential between electrons is obtained as

$$V_{\mathrm{eff}}(\boldsymbol{q}) = \left|g_q^{ads}\right|^2 \cdot \frac{2\hbar\omega_{\mathrm{ads}}}{(\hbar\omega)^2 - (\hbar\omega_{\mathrm{ads}})^2 + i\delta\cdot\mathrm{sgn}(\omega)} \quad (5)$$

where the adsorption mode with momentum $\boldsymbol{q}$ and frequency $\omega_{\mathrm{ads}}$ exchanged in the middle, $\hbar\omega = \varepsilon_{\boldsymbol{k}+\boldsymbol{q}} - \varepsilon_{\boldsymbol{k}}$ is the energy difference of electronic transition, $i\delta$ in the denominator ensures causality (delayed response).

More importantly, different from the phonon-mediated dynamic resonance of BCS, the electronic pairing here is a selective condensation driven by the chemical potential equilibrium. Only electrons with energies within the range of $\pm\varepsilon_p$ near the Fermi surface can be captured by the adsorption potential on the lattice surface into the adsorption space, thereby participating in Cooper pairing. Electrons with energies outside this range do not "feel" the adsorption potential and maintain normal state behavior. In this case Eq.(5) can be revised to

$$V_{\mathrm{eff}}(\boldsymbol{q}) = \left|g_q^{ads}\right|^2 \cdot \frac{2\hbar\omega_{\mathrm{ads}}}{(\hbar\omega)^2 - (\hbar\omega_{\mathrm{ads}})^2 + i\delta\cdot\mathrm{sgn}(\omega)} \cdot \Theta\left(\varepsilon_p - |\varepsilon_{\boldsymbol{k}} - \mu|\right) \quad (6)$$

where $\Theta$ is the Heaviside function, which means that electrons can only pair when their energy is within the range $\varepsilon_p$ near the chemical potential $\mu$.

In addition, near the Fermi surface where superconductive pairing occurs, the electronic energy transfer satisfies $\hbar\omega = \varepsilon_{\boldsymbol{k}+\boldsymbol{q}} - \varepsilon_{\boldsymbol{k}} \ll \hbar\omega_{\mathrm{ads}}$, so the real part of the denominator becomes negative, and the retarded interaction automatically becomes attractive.

In the adsorption space near the lattice surface, the electron gas is subject to interface constraints, producing Friedel oscillations, with the electron density exhibiting a periodic distribution along the direction perpendicular to the surface characterized by a wave vector of $2k_F$ ($k_F$ is Fermi wave vector). High-temperature superconductors generally have a layered Fermi surface structure, which readily exhibits significant Fermi surface nesting effects. When the nesting wave vector is close to resonance with the characteristic wave vector of Friedel oscillations, the electronic polarizability is greatly enhanced, and the collective electron-hole response is significantly amplified. This synergy strengthens the local binding of electrons by the adsorption potential, increases the oscillation frequency and coherence strength of the adsorbed modes, and further enhances the effective electron-electron attractive interaction, promoting Cooper pair condensation and the emergence of a higher superconducting transition temperature Tc.

Because of this Thomas-Fermi screening in the electron gas, the response of the chemical potential needs to be divided by the dielectric function, i.e. $\frac{\partial\mu}{\partial n} \rightarrow \frac{\partial\mu}{\partial n} \cdot \frac{q^2}{q^2+k_{\mathrm{TF}}{}^2}$, where Thomas-Fermi wavenumber $k_{\mathrm{TF}} = \sqrt{\frac{3e^2\mathrm{n}}{2\varepsilon_0\varepsilon_r\varepsilon_F}}$ by using universal symbols. According to our previous work [25], in the superconducting phase there is

$$\frac{\partial\mu}{\partial n_s} = 3.19\frac{k_B\mathrm{T}}{\omega^2 n}\left(\frac{m^* k_B T}{\hbar^2 n^{2/3}}\right)^{9/11} \tag{7}$$

where $k_B$ is the Boltzmann constant, $T$ is the temperature, and $\omega = n_s/n$ is the relative proportion of the paired electrons.

Therefore, the electron-adsorption-mode coupling strength can be further improved to

$$g_q^{ads} = \frac{\partial\mu}{\partial n_s}\sqrt{\frac{n_s}{V}} \cdot \frac{q^2}{q^2+k_{\mathrm{TF}}{}^2} = 3.19\frac{k_B\mathrm{T}}{\omega\sqrt{\omega n V}}\left(\frac{m^* k_B T}{\hbar^2 n^{2/3}}\right)^{9/11} \cdot \frac{q^2}{q^2+k_{\mathrm{TF}}{}^2} \tag{8}$$

Further the effective interaction potential between electrons is

$$|V_{\mathrm{eff}}(\boldsymbol{q})| \approx \frac{2|g_q^{ads}|^2}{\hbar\omega_{\mathrm{ads}}}, \tag{9}$$

and by analogy with the electron-phonon coupling constant, we define the adsorption coupling constant as

$$\lambda_{ads} = N(0)|V_{\mathrm{eff}}(\boldsymbol{q})| = \frac{m^* k_F}{2\pi^2\hbar^2}|V_{\mathrm{eff}}(\boldsymbol{q})| = \frac{m^* k_F \cdot |g_q^{ads}|^2}{\pi^2\hbar^3\omega_{\mathrm{ads}}} \tag{10}$$

where $N(0)$ is Fermi surface single-spin state density, and $k_F$ is Fermi wavenumber.

For $YBa_2Cu_3O_7$, substituting $\omega$=0.0306, $T$c=90K, $n$=3.0×10$^{27}$, $m^*=3m$, $V = 173.28\text{Å}^3$ that is the unit cell volume, $\hbar\omega_{\mathrm{ads}} \sim \varepsilon_p = 3m\mathrm{eV}$, and $k_F = 0.6\text{Å}^{-1}$ into Eq.(10), we obtain that $\lambda_{ads} \approx 43.4$, which means that $YBa_2Cu_3O_7$ is strong-coupling superconductor.

## 4. *d*-wave symmetry and Coherence length

Performing a Fourier transform of Eq.(6) to real space, then there is

$$V_{\text{eff}}(\boldsymbol{r}) = \frac{\left|g_q^{ads}\right|^2 \omega_{\text{ads}}}{(2\pi)^3} \int d^3q \frac{e^{i\boldsymbol{q}\cdot\boldsymbol{r}}}{\left(\hbar v_{\text{F}} q\right)^2 - \omega_{\text{ads}}{}^2} \cdot \Theta\left(\varepsilon_p - \hbar v_{\text{F}} q\right)$$

$$\approx -\frac{\left|g_q^{ads}\right|^2}{\omega_{\text{ads}}\left(\hbar v_{\text{F}}\right)^2} \cdot \frac{e^{-r/(\hbar v_{\text{F}}/\varepsilon_p)}}{r} \cdot \cos(2k_{\text{F}} r) \quad (11)$$

where $v_{\text{F}}$ is Fermi velocity, and $\cos(2k_{\text{F}} r)$ is caused by Friedel oscillations (Fermi surface nesting effect).

### 4.1 The origin of d-wave symmetry

In the $CuO_2$ plane, the layered anisotropy of the adsorption potential and the fourfold symmetry of the oxygen sites lead to a momentum dependence of the coupling constant $g_q^{ads}$, i.e.

$$\left|g_q^{ads}\right|^2 \propto \left(\cos q_x a - \cos q_y b\right)^2 \quad (12)$$

where *a* and *b* are the lattice constant in the *a* direction and *b* direction of copper oxide superconductors, respectively.

Substituting Eq.(12) into Eq.(6), we have

$$V_{\text{eff}}(\boldsymbol{q}) \propto \left[\cos(k_x - k_x')a - \cos(k_y - k_y')b\right]^2 \cdot \frac{2\hbar\omega_{\text{ads}}}{\left(\varepsilon_{\boldsymbol{k}} - \varepsilon_{\boldsymbol{k}}'\right)^2 - \left(\hbar\omega_{\text{ads}}\right)^2} \quad (13)$$

Near the Fermi surface, $\varepsilon_{\boldsymbol{k}} \approx \varepsilon_{\boldsymbol{k}}' \approx \mu$, thus

$$V_{\text{eff}}(\boldsymbol{q}) \propto \left[\cos(k_x a) - \cos(k_y b)\right]\left[\cos(k_x' a) - \cos(k_y' b\right] \quad (14)$$

which is exactly the factored form of the $d_{x^2-y^2}$ wave pairing.

This is the fundamental difference from BCS theory, which shows that the *d*-wave does not come from strong correlation repulsion (such as in RVB theory [31]), but from the lattice symmetry of the adsorption potential and the anisotropy of interfacial adsorption forces.

### 4.2 Coherence length determined by the effective interaction potential $V_{\text{eff}}(\boldsymbol{q})$

Considering the effective interaction potential $V_{\text{eff}}(\boldsymbol{q})$ additionally induced by the adsorption potential, the one-dimensional Ginzburg-Landau equation in the absence of a magnetic field could be rewritten as

$$-\frac{\hbar^2}{2m^*}\frac{d^2\varphi}{dx^2} + (\alpha + V_{\text{eff}})\varphi + \beta|\varphi|^2\varphi = 0 \quad (15)$$

Let $f = \varphi - \varphi_0$ ( $\Phi_0$ is the rigid effective wave function with $|\varphi_0|^2 = -\alpha/\beta$), then there is

$$\frac{d^2 f}{dx^2} - \frac{4m^*(|\alpha| + V_{\text{eff}})}{\hbar^2} f = 0 \quad (16)$$

From Eq.(16) the coherence length can be obtained as

$$\xi = \sqrt{\frac{\hbar^2}{4m^*(|\alpha| + V_{\text{eff}})}} \sim \frac{\hbar}{2\sqrt{m^* V_{\text{eff}}}} \quad (17)$$

In addition, through polarization-dependent ultrafast electron diffraction experiments, the momentum-space anisotropy of electron-phonon coupling in YBCO was directly measured [32]. The electron-phonon coupling constant $\lambda$=0.55 is the strongest in the Cu-O bond direction with the largest energy gap, while $\lambda$=0.08 is the weakest in the node direction, and the average coupling constant in all directions $\lambda$=0.26. Thus the directionality of the potential $V_{\text{eff}}(\boldsymbol{q})$ in Eq.(13) can be determined by the ratio of these coupling constants.

For $YBa_2Cu_3O_7$, considering the anisotropy of interfacial adsorption forces, in out-of-plane (*c*-axis) by taking $V_{\text{eff}} \cdot \frac{0.55}{0.08}$ into Eq.(17) we have $\xi \approx 1.34\text{nm}$, while in the *a-b* plane by taking $V_{\text{eff}} \cdot \frac{0.55}{0.26}$ into Eq.(17) $\xi \approx 0.35\text{nm}$, which are very close to the literature results [33-34].

In addition, the effective superconducting component ratio $\omega \approx 0.0306$ in $YBa_2Cu_3O_7$ [25] is closely related to the short coherence length $\xi$ inherent to cuprate superconductors. The coherence length $\xi$ is an intrinsic quantum property determined by the pairing interaction and Fermi velocity, rather than the carrier density. The small value of $\omega$ arises because the short $\xi$ restricts the spatial range of Cooper-pair coherence and allows only a small fraction of carriers to form the coherent superconducting condensate. Within the adsorption potential theory, the layer-dependent adsorption potential further selects and localizes superconducting carriers within the $CuO_2$ planes, leading to the observed low $\omega$.

### 4.3 Pseudo-energy gap Behavior

According to the adsorption potential theory [25], at $T = T_c$ the adsorption potential $\varepsilon_p > 0$ and Heaviside truncation in Eq.(6) is still present, despite the disappearance of the energy gap of the BCS theory. In this case electrons are still 'pre-paired' by the adsorption potential (nonzero amplitude) within the $\pm\varepsilon_p$ window, but thermal fluctuations destroy phase rigidity, so there is no superconductivity and the bandgap-like feature (pseudo-bandgap) persists until $\varepsilon_p = 0$.

### 4.4 Superconducting gap verified by STM/S measurement

Our previous results reveal that those high-$T_c$ superconductors ($T_c \geq 40\text{K}$) are mainly formed by the molar adsorption potentials, and the low-$T_c$ superconductors ($T_c < 40\text{K}$) by both the molar adsorption potentials and the energy gap of the BCS theory. Based on this, the BCS energy gap equation is corrected to

$$\Delta(T) = N(0)|V_{\text{eff}}(\boldsymbol{q})| \int_0^{\hbar\omega_{\text{D}}+\varepsilon_{\text{P}}} d\xi \frac{\Delta}{\sqrt{\xi^2+\Delta^2}} \tanh\left(\frac{\sqrt{\xi^2+\Delta^2}}{2k_B T}\right) \tag{18}$$

Because for high-temperature superconductor, $\varepsilon_p \sim 2-4\text{meV} \ll \hbar\omega_{\text{D}} \sim 50\text{meV}$, Eq.(18) can be further simplified as

$$\Delta(T=0) \approx N(0)|V_{\text{eff}}(\boldsymbol{q})| \left[\ln \frac{\hbar\omega_{\text{D}}}{2k_B T_c} - \left(\gamma + \ln\frac{4}{\pi}\right) + \frac{\varepsilon_p}{\hbar\omega_{\text{D}}}\right] \tag{19}$$

where $\gamma = 0.577$ is Euler's constant, and $\omega_{\text{D}}$ is Debye frequency.

For $YBa_2Cu_3O_7$, substituting $N(0)|V_{\text{eff}}(\boldsymbol{q})|$=43.4, $\hbar\omega_{\text{D}} = 50\text{meV}$, $T_c = 90\text{K}$, and $\varepsilon_p = 3\text{meV}$ into Eq.(19), we obtain that the superconducting gap $\Delta = 18.37\text{meV}$, which is very close to the measured result of 17meV by Scanning Tunneling Microscopy/ Spectroscopy [35].

## 5. Conclusion

In this paper, the electron pairing mechanism by the interfacial charge-induced adsorption mode of high-temperature superconductors is revealed. The effective interaction potential between electrons is exactly derived in details, from which we verify that *d*-wave come from the anisotropy of interfacial adsorption forces, and explain the pseudo-energy gap behavior, and further establish the energy gap equation. By using the one-dimensional Ginzburg-Landau equation in the absence of a magnetic field, we obtain the coherence length expression. For the YBCO superconductors, the electron-adsorption-mode coupling constant is up to 43.4, the coherence length calculated is very close to the literature results, and further we obtain the superconducting gap $\Delta = 18.37\text{meV}$, which is very close to the measured result of 17meV by Scanning Tunneling Microscopy/Spectroscopy. These quantitative predictions close to the known results could verify our theoretical framework.